\documentclass[journal]{IEEEtran}
\usepackage{amssymb}
\usepackage{amsmath}

\usepackage{color}
\usepackage{cite}

\definecolor{mygreen}{rgb}{0, 0.4, 0}
\definecolor{mypur}{rgb}{0.4, 0, 0.4}

\ifCLASSINFOpdf
 \usepackage[pdftex]{graphicx}
\else
\fi
\usepackage{array}
\usepackage[caption=false,font=footnotesize]{subfig}
\usepackage{dblfloatfix}

\ifCLASSOPTIONcaptionsoff
  \usepackage[nomarkers]{endfloat}
 \let\MYoriglatexcaption\caption
 \renewcommand{\caption}[2][\relax]{\MYoriglatexcaption[#2]{#2}}
\fi
\usepackage{url}

\begin{document}

\title{Absolute Eigenvalues-Based Covariance Matrix Estimation for a Sparse Array}

\author{Kaushallya Adhikari
                       \thanks{This work was supported by the U.S. Office of Naval Research under Grant N00014-20-1-2820. The author is with the University of Rhode Island. 
                          e-mail: kadhikari@uri.edu}}

\maketitle

\sloppy

\begin{abstract}
The ensemble covariance matrix of a wide sense stationary signal spatially sampled by a full linear array is positive semi-definite and Toeplitz. However, the direct augmented covariance matrix of an augmentable sparse array is Toeplitz but not positive semi-definite, resulting in negative eigenvalues that pose inherent challenges in its applications, including model order estimation and source localization. The positive eigenvalues-based covariance matrix for augmentable sparse arrays is robust but the matrix is unobtainable when all noise eigenvalues of the direct augmented matrix are negative, which is a possible case. To address this problem, we propose a robust covariance matrix for augmentable sparse arrays that leverages both positive and negative noise eigenvalues. The proposed covariance matrix estimate can be used in conjunction with subspace based algorithms and adaptive beamformers to yield accurate signal direction estimates.
\end{abstract}



\section{Introduction}

Estimation of directions of arrival (DOAs) of wide-sense stationary (WSS) signals using the data sampled by an array of sensors is an important task in many fields such as sonar, radar, seismology, radio astronomy, and neurophysiology. The data sampled by an $L$-element uniform linear array (ULA) that has an intersensor spacing of $\lambda/2,$ where $\lambda$ is the wavelength of the WSS signals impinging on the array, can be used to resolve up to $L-1$ DOAs \cite{VanTrees,JD}. There exist augmentable sparse arrays for which the number of resolvable DOAs exceeds the number of sensors. An $L_S$-element augmentable sparse array that has an aperture of $(L_F-1)\lambda/2$ can resolve $L_A-1$ DOAs where $L_F\geq L_A\geq L_S$. Sparse arrays where $L_A=L_F$ are called fully augmentable arrays and sparse arrays where $L_S<L_A<L_F$ are called partially augmentable arrays.  Minimum redundant arrays \cite{Moffet} and nested arrays \cite{nested1} are examples of fully augmentable sparse arrays. Semi-coprime arrays \cite{adhikariJasa1} are examples of partially augmentable sparse arrays. Coprime arrays \cite{VandP1} may be partially or fully augmentable based on the design parameters. 

The accuracy and resolution of DOA estimates for any array also depends on the estimation algorithm. Conventional beamforming-based algorithms such as product processing and min processing have lower resolution than algorithms that are based on estimates of second order statistics (signal covariance matrix) such as multiple signal classification (MUSIC) \cite{Schmidt} or minimum variance distortionless response (MVDR) beamformer \cite{shadings,MartinoIodice,AdhikariAccess1,liubuck4,detection2019,KBW,chavali2,icasspdetection,Rooney1} for augmentable sparse arrays. This work focuses on the algorithm that uses signal covariance estimates.

The performance of a DOA estimation algorithm such as MUSIC and MVDR depends on the accuracy of the covariance matrix estimate. The ensemble covariance matrix associated with an impinging WSS signal possesses two important properties: positive semi-definite and Toeplitz structure. Many covariance matrix estimates found in the literature do not possess one or both of these properties. For instance, the sample covariance matrix (SCM) of a full ULA (to be defined in Section~\ref{Sec:model}), which is widely used in DOA estimation, is positive semi-definite but is not guaranteed to be Toeplitz. The direct augmented matrix (DAM) of a sparse array (to be defined in Section~\ref{Sec:model}) is Toeplitz but is not guaranteed to be semi-definite \cite{Abramovich,psdestimation}. For sparse arrays, Abramovich \emph{et al.} have presented an algorithm for estimating a positive definite Toeplitz covariance matrix, called positive eigenvalues-based matrix (PEM) \cite{Abramovich}. This algorithm generates reasonable covariance matrix estimates when the corresponding DAM  has at least one non-negative noise eigenvalue. However, when the DAM's noise eigenvalues are all negative, this algorithm is inapplicable. We present a new algorithm to estimate covariance matrix for sparse arrays that works even when all noise eigenvalues of the DAM are negative. Morever,  when the proposed covariance matrix is used in conjunction with subspace based methods such as MUSIC \cite{Schmidt} or adaptive algorithms such as MVDR beamformer \cite{Capon}, the DOA estimates are more accurate or commensurate with the PEM.

\section{Problem Formulation}
\label{Sec:model}

Consider an $L_S$-sensor linear array oriented along the positive $z$-axis. The sensor locations along the $z$-axis are $d_1=0,$ $d_2\lambda/2$, $d_3\lambda/2$, $...$, $d_{L_S}\lambda/2=(L_F-1)\lambda/2$, where $d_2$, $d_3$, $...$, $d_{L_S}$ are assumed to be positive integers. Note that a ULA is a special case of this $L_S$-sensor array where $d_2=1$, $d_3=2$, $...$, $d_{L_S}=L_S-1.$ If the data sampled by the array consist of $Q$ WSS uncorrelated planewaves in white Gaussian noise, the received data vector at time $t$ is given by
\begin{equation}
\label{modeleq}
\mathbf{x}(t)=\sum_{i=1}^Q\mathbf{v}(\theta_i)s_i(t) + \mathbf{n}(t),
\end{equation}
where $s_i(t)$ is the complex amplitude of the $i^{th}$ signal, $\mathbf{v}(\theta_i)$ is the array manifold vector corresponding to the $i^{th}$ signal, $\theta_i$ is the $i^{th}$ planewave angle measured from the positive $z$-axis, and $\mathbf{n}(t)$ is the $L_S$-element noise vector. We assume that the variables $s_i(t)$ and $\mathbf{n}(t)$  are zero-mean proper complex Gaussian random variable and random vector, respectively. The $k^{th}$ element of the array manifold vector $\mathbf{v}(\theta_i)$ is $\exp(j\pi u_id_k),$ where $u_i=\cos(\theta_i)$ is the direction cosine.

The SCM corresponding to the data vector in (\ref{modeleq}) is $\mathbf{x}(t)\mathbf{x}(t)^H$. Since the SCM is the outer product of a vector with itself, it is guaranteed to be positive semi-definite. If there are $T$ snapshots sampled at times $t=t_1,$ $t_2,$ $...,$ $t_T,$ respectively, the SCM is given by the average of the data vectors' outer products, $\boldsymbol{\mathcal{S}}=\sum_{k=1}^T\mathbf{x}(t_k)\mathbf{x}^H(t_k)/T$. The  dimension of $\boldsymbol{\mathcal{S}}$ is $L_S\times L_S.$ To obtain a Toeplitz estimate, $\boldsymbol{\mathcal{T}}$, we can apply redundancy averaging, which entails replacing every element of $\boldsymbol{\mathcal{S}}$ by the average of the elements along the diagonal in which it resides. Since the dimension of $\boldsymbol{\mathcal{T}}$ is $L_S\times L_S$, it can be used to estimate up to only $L_S-1$  DOAs. Sparse arrays such as MRAs, coprime arrays, and nested arrays are popular for their ability to resolve more DOAs than the number of sensors, which cannot be realized with $\boldsymbol{\mathcal{S}}$ or $\boldsymbol{\mathcal{T}}.$ To utilize all degrees of freedom offered by a sparse array, consider the direct augmented matrix (DAM) which is obtained as described below:

\begin{itemize}[label=,leftmargin=*]
\item \textbf{Step 1:} Find the possible covariance estimate at each possible lag as $r_{(0)}[k]=y[k]\star y^*[-k]$, where the symbol $\star$ denotes discrete convolution and $y[k]$ is the measurement made by the sensor at $z=k\lambda/2.$ If there is no sensor present at $z=k\lambda/2,$ $y[k]$ is considered $0.$ The  covariance function $r_{(0)}[k]$ is defined for lags in the range  $-(L_F-1)\leq k\leq L_F-1.$ However, some of the values of $r_{(0)}[k]$ are $0$ in the range $-(L_F-1)\leq k\leq L_F-1$ if the sparse array is only partially augmentable.
\item  \textbf{Step 2:} Extract the hole-free covariance function estimate, $r_{(1)}[k]$ by truncating $r_{(0)}[k]$. This step is not needed in a fully augmentable sparse array. The function $r_{(1)}[k]$ is defined for lags in the range $-(L_A-1)\leq k \leq L_A-1.$
\item  \textbf{Step 3:} Remove the redundancy in $r_{(1)}[k]$ by discarding the values corresponding to negative lags. We will refer to the new covariance function as $r_{(2)}[k]$, which is defined for lags in the range $0\leq k\leq L_A-1.$
\item  \textbf{Step 4:} Remove the bias in $r_{(2)}[k]$ to obtain $r_{(3)}[k]$ as $r_{(2)}[k]/ c[k]$, where $c[k]$ is the hole-free coarray of the sparse array corresponding to positive lags.
\item  \textbf{Step 5:} The DAM is obtained by forming a Toeplitz matrix that has the elements  $r_{(3)}[k]$ for $k=0,$ $1,$ $...,$ $L_A-1$ along its first row. This matrix is denoted by $\boldsymbol{\mathcal{T}}_{DAM}$ in the sequel.
\end{itemize}

The DAM can be directly used with eigenanalysis based DOA estimation algorithms such as MUSIC \cite{AdhikariAccess2,AdhikariNAECON}. However, the DAM is not guaranteed to be positive semi-definite \cite{Abramovich,psdestimation}.

\section{Positive Eigenvalues-Based Matrix}
\label{sec:pem}

An iterative algorithm to estimate a positive definite covariance matrix of dimension $L_A\times L_A$ for a sparse array is described in \cite{Abramovich}. This algorithm uses the alternating convex projection method  in \cite{Grigoriadis} to compute an augmented covariance matrix that is both Toeplitz and positive definite. The algorithm keeps alternating between enforcing Toeplitz property and positive definiteness until the algorithm converges to a matrix that is both Toeplitz and positive definite. We summarize the essential components of their algorithm below:
\begin{itemize}[label=,leftmargin=*]
\item  \textbf{Step 1:} Set $\boldsymbol{\mathcal{T}}_{DAM}$ as the initial estimate of the covariance matrix: $\hat{\mathbf{T}}^{(0)}=\boldsymbol{\mathcal{T}}_{DAM}$.
\item  \textbf{Step 2:} If $\hat{\lambda}_1,$ $\hat{\lambda}_2,$ $...,$ $\hat{\lambda}_{L_A}$ are  the eigenvalues of $\hat{\mathbf{T}}^{(0)}$ in descending order, form the positive definite Hermitian matrix $\hat{\mathbf{H}}$ corresponding to $\hat{\mathbf{T}}^{(0)}$ as
\begin{equation}
\hat{\mathbf{H}}=\bar{\lambda}_n\mathbf{I}_{L_A}+\sum_{j=1}^Q\Bigl(\hat{\lambda}_j-\bar{\lambda}_n\Bigr)\mathbf{v}_j\mathbf{v}_j^H,
\end{equation}
where
\begin{equation}
\label{avgeig}
\bar{\lambda}_n=\frac{1}{L_A-Q}\sum_{j=Q+1}^{L_A}\hat{\lambda}_j
\end{equation}
 is the average of the noise eigenvalues and $\mathbf{I}_{L_A}$ is the $L_A\times L_A$ identity matrix.
\item  \textbf{Step 3:} Given the positive definite matrix $\hat{\mathbf{H}}$, form the Toeplitz matrix $\hat{\mathbf{T}}^{(1)}$ by applying redundancy averaging.
\item  \textbf{Step 4:} Let $\hat{\lambda}_1,$ $\hat{\lambda}_2,$ $...,$ $\hat{\lambda}_{L_A}$ be  the eigenvalues of $\hat{\mathbf{T}}^{(1)}$ in descending order. If $\frac{\hat{\lambda}_{Q+1}-\hat{\lambda}_{L_A}}{\hat{\lambda}_{L_A}}\geq \epsilon <<1,$ where $\epsilon$ is a small positive number, then replace $\hat{\mathbf{T}}^{(0)}$ in \textbf{Step 1} by $\hat{\mathbf{T}}^{(1)}$ and repeat \textbf{Step 2}, \textbf{3}, and \textbf{4}.
\end{itemize}
This algorithm converges when the largest noise eigenvalue is approximately equal to the smallest noise eigenvalue. \cite{Abramovich} mentions that some of the noise eigenvalues of the DAM might be negative. To tackle this problem, \cite{Abramovich} suggests discarding the negative noise eigenvalues and modifying (\ref{avgeig}) as
\begin{equation}
\label{avgeigpos}
\bar{\lambda}_n=\frac{1}{\sum_{j=Q+1}^{L_A}I[j]}\hspace*{0.2cm}\sum_{j=Q+1}^{L_A}\hat{\lambda}_jI[j], 
\end{equation}
where $I[j]=1$ if $\hat{\lambda}_j\geq 0$ and $I[j]=1$ if $\hat{\lambda}_j< 0.$

\subsection{Degenerative Case for PEM}
\label{sec:degen}

\begin{figure}[ht]
\centerline{\includegraphics[width=0.8\columnwidth, trim = 0cm 21.75cm 0cm 0cm]{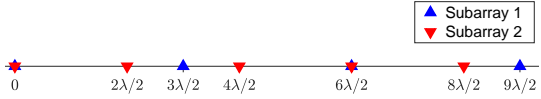}}
\caption{A fully augmentable coprime array formed by interleaving Subarray 1 that has $4$ sensors and an intersensor spacing of $3\lambda/2$ and Subarray 2 that has $5$ sensors and an intersensor spacing of $2\lambda/2$. }
\label{coprimefiglabel}
\end{figure}

One important possibility that has not been addressed in \cite{Abramovich} is that every noise eigenvalue of the DAM of a sparse matrix may be negative. The algorithm in \cite{Abramovich} is not applicable when all noise eigenvalues are negative because it relies on having a positive subset of the noise eigenvalues to obtain $\bar{\lambda}_n$ in (\ref{avgeigpos}). To illustrate this point, we consider a fully augmentable sparse array depicted in Fig.~\ref{coprimefiglabel}. It is a coprime array formed by interleaving Subarray~1 that has $4$ sensors and an intersensor spacing of $3\lambda/2$ and Subarray~2 that has $5$ sensors and an intersensor spacing of $2\lambda/2$. The number of sensors in the array is $7$ and the aperture is $9\lambda/2.$ The corresponding DAM is of dimension $10\times 10.$ Consider a scenario with two planewave sources, each of SNR $25$ \rm{dB}. The source direction cosines are selected as $u_1=0.2165\frac{4}{10}$ and $u_2=-0.2165\frac{4}{10}$ so that the source locations are half-power beamwidth apart for a full ULA with $9\lambda/2$ aperture \cite[p. 999]{VanTrees}. We used $25$ snapshots to compute the DAM and find the corresponding eigenvalues. Note that the number of negative eigenvalues differ from trial to trial. The eigenvalues of the DAM were found to be $2603.9$, $294.9$, $-10.2$, $-6.9$, $-6$, $-3.7$, $-3.1$, $-2.1$, $-1.7$, and $-1.2$ in one of the trials. The first two eigenvalues are signal plus noise eigenvalues and are positive. The remaining $8$ eigenvalues are noise eigenvalues and all values are negative. A  set of all negative noise eigenvalues poses an inherent challenge in covariance matrix estimation. In the next section, we propose a covariance matrix estimate to address this challenge.

\section{Absolute Eigenvalues-Based Matrix}
\label{sec:aem}

\begin{figure}
\centerline{\includegraphics[width=0.9\columnwidth, trim = 0cm 18cm 0cm 0cm]{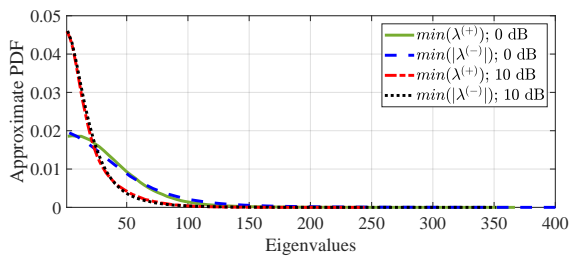}}
\caption{Comparison of normalized histograms (approximated PDFs) of the minimum positive noise eigenvalue ($min(\hat{\lambda}^{(+)})$) and the minimum absolute value of negative noise eigenvalue ($min(|\hat{\lambda}^{(-)}|)$) at two different SNR values.
}
\label{eigpdflabel}
\end{figure}

When the noise eigenvalues of the DAM are all negative, it renders the alternating convex projection method of covariance estimation inapplicable. We propose a method to estimate the augmented covariance matrix that leverages negative noise eigenvalues instead of discarding them. The proposed covariance matrix estimate is referred to as absolute eigenvalues-based matrix (AEM). To motivate utilization of negative eigenvalues, consider the coprime array in Fig.~\ref{coprimefiglabel} in the two planewave scenario described in Section~\ref{sec:degen}. We approximated the probability density functions (PDFs) of two random variables: (1) minimum positive eigenvalue, $min(\lambda^{(+)})$, of the DAM; and (2) minimum of the absolute values of the negative  eigenvalues, $min(|\lambda^{(-)}|)$. These PDFs were obtained using $1,000,000$ realizations of the DAM, with each realization consisting of $10$ snapshots. The PDFs are depicted in Fig.~\ref{eigpdflabel} for $0$ \rm{dB} and  $10$ \rm{dB} SNR. The close resemblance between the PDFs of $min(\lambda^{(+)})$ and $min(|\lambda^{(-)}|)$ suggests that negative noise eigenvalues can be used in improving the value of $\bar{\lambda}_n$. Therefore, given the eigenvalues of the DAM in descending order of their magnitudes $|\hat{\lambda}_1|\geq|\hat{\lambda}_2|\geq, ...,\geq |\hat{\lambda}_{L_A}|$, we propose calculating average noise eigenvalues as
\begin{equation}
\label{avgeigabs}
\bar{\lambda}_n=\frac{1}{L_A-Q}\sum_{j=Q+1}^{L_A}|\hat{\lambda}_j|.
\end{equation}
The AEM is then formed as
\begin{equation}
\label{aemeq}
\boldsymbol{\mathcal{R}}=\sum_{j=1}^Q\hat{\lambda}_j\mathbf{v}_j\mathbf{v}_j^H+\sum_{j=Q+1}^{L_A}\bar{\lambda}_n\mathbf{v}_j\mathbf{v}_j^H,
\end{equation}
where $\mathbf{v}_j$ is the eigenvector corresponding to $|\lambda_j|$. Since each term in the two sums of (\ref{aemeq}) is a vector's outer product with itself, scaled by a non-negative scalar, $\boldsymbol{\mathcal{R}}$ is guaranteed to be positive semi-definite. This approach does not fail when all noise eigenvalues are negative. Moreover, when only some of the noise eigenvalues are negative, using (\ref{avgeigabs}) provides an estimate of $\bar{\lambda}_n$ that has less variance since it is obtained by averaging over more realizations compared to (\ref{avgeigpos}). Note that $\boldsymbol{\mathcal{R}}$ in (\ref{aemeq}) is not guaranteed to be Toeplitz. We can alternate between Toeplitz and positive semi-definite estimates following the approach used in the PEM computation. However, our simulation results in Section~\ref{Sec:results} show that $\boldsymbol{\mathcal{R}}$ in (\ref{aemeq}) is robust and does not warrant further improvement with an iterative method.

\section{Results}
\label{Sec:results}

\begin{figure}[ht]
\centerline{\includegraphics[width=0.95\columnwidth, trim = 0cm 13cm 0cm 2cm]{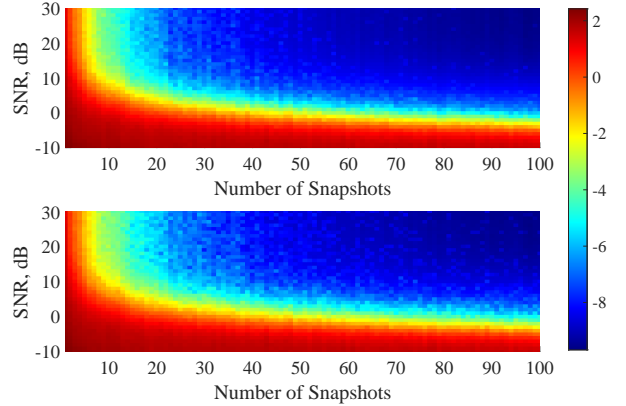}}
\caption{Comparison of RMSEs for MUSIC using PEM (top panel) and AEM (bottom panel).}
\label{music3dlabel}
\end{figure}
\begin{figure}[ht]
\centerline{\includegraphics[width=0.95\columnwidth, trim = 0cm 13cm 0cm 2cm]{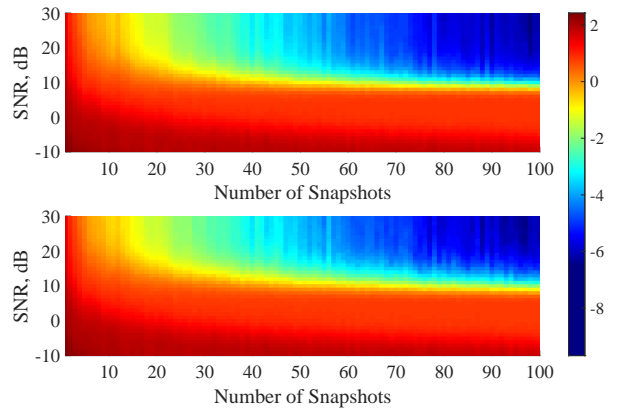}}
\caption{Comparison of RMSEs for MVDR using PEM (top panel) and AEM (bottom panel).}
\label{mvdr3dlabel}
\end{figure}
\begin{figure}[ht]
\centerline{\includegraphics[width=0.95\columnwidth, trim = 0cm 13cm 0cm 1cm]{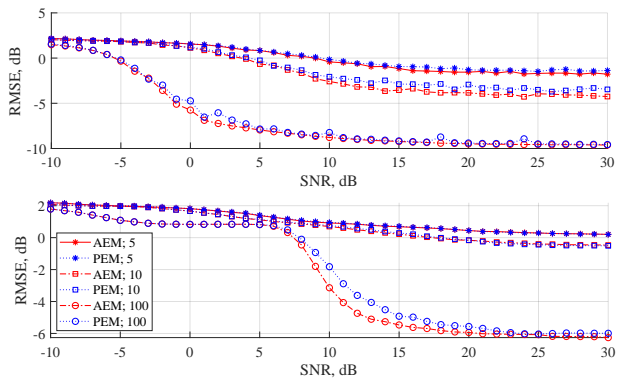}}
\caption{Comparison of RMSEs in DOA for MUSIC (top panel) and MVDR (bottom panel) when AEM (red lines) and PEM (blue lines) are used as covariance matrix estimates.}
\label{rmsecutsfiglabel}
\end{figure}

This section compares the DOA estimation performance of MUSIC and MVDR using the covariance matrices PEM and AEM. We use the array in Fig.~\ref{coprimefiglabel} in the two uncorrelated planewave scenario described in Section~\ref{sec:degen}. For fair comparison between the PEM and AEM based DOA estimation, we used the same $1000$ datasets, discarding any dataset that consisted of all negative noise eigenvalues since the PEM cannot be evaluated for such datasets.

Fig.~\ref{music3dlabel} compares the root mean squared error (RMSE) in \rm{dB} scale for DOA estimation with MUSIC using PEM and AEM over a range of SNRs and numbers of snapshots. The top panel corresponds to the PEM-based MUSIC and the bottom panel corresponds to the AEM-based MUSIC. Comparing the two panels, we can infer that there is negligible difference between AEM and PEM based MUSIC, in cases when PEM is not degenerative. Similarly, Fig.~\ref{mvdr3dlabel} compares the RMSE with MVDR using PEM (top panel) and AEM (bottom panel). For MVDR also, there is no noticeable difference in the two panels. In Fig.~\ref{rmsecutsfiglabel}, we plot several cuts of Fig.~\ref{music3dlabel} and Fig.~\ref{mvdr3dlabel}. We consider three different numbers of snapshots: $5$ (less than the number of sensors), $10$ (equal to the number of sensors), and $100$ (greater than the number of sensors). The solid linestyle with asterisk marker indicates $5$ snapshots, the dash-dot linestyle with square markers indicates $10$ snapshots, the dotted line with circle marker indicates $100$ snapshots; and the red and blue colors indicate AEM and PEM, respectively. From these RMSE plots, we can infer that in the cases when PEM is evaluable, the AEM-based estimation is either as accurate as or more accurate than the PEM-based estimation. This is true for both MUSIC and MVDR, which demonstrates that AEM is more robust than PEM.

\section{Conclusion}

We presented a robust positive semi-definite covariance matrix, AEM, for  sparse arrays. The proposed approach does not degenerate when all eigenvalues of the DAM are negative. As evidenced by the simulations, the AEM is more robust than the competing covariance matrix estimate, PEM.

\ifCLASSOPTIONcaptionsoff
  \newpage
\fi

\bibliographystyle{IEEEtran}

\bibliography{IEEEabrv,ToeplitzSPL}

\end{document}